\documentclass[useAMS]{mn2e}
 \usepackage{times}
 \usepackage{graphicx}
 \usepackage{epsfig}
 \usepackage{amssymb}

 \title[Sudden Earth impacts: are they truly random?]
       {Recent multi-kiloton impact events: are they truly random?}

 \author[C. de la Fuente Marcos and R. de la Fuente Marcos]
        {C.~de~la~Fuente~Marcos\thanks{E-mail: nbplanet@fis.ucm.es}
         and
         R. de la Fuente Marcos \\
         Universidad Complutense de Madrid,
         Ciudad Universitaria, E-28040 Madrid, Spain}
 \date{Accepted 2014 August 29.
       Received 2014 August 28;
       in original form 2014 August 11}
 \pubyear{2014}
 
\begin{document}
  \maketitle

  \begin{abstract}
     It is customarily assumed that Earth-striking meteoroids are completely 
     random, and that all the impacts must be interpreted as uncorrelated 
     events distributed according to Poisson statistics. If this is correct, 
     their impact dates must be uniformly spread throughout the year and 
     their impact coordinates must be evenly scattered on the surface of our 
     planet. Here, we use a time- and yield-limited sample of Earth-impacting 
     superbolides detected since 2000 to explore statistically this critical 
     though frequently overlooked topic. We show that the cadence of these 
     multi-kiloton impact events is incompatible with a random fall pattern 
     at the 0.05 significance level or better. This result is statistically 
     robust and consistent with the observed distribution of the longitudes 
     of the ascending nodes of near-Earth objects (NEOs). This lack of 
     randomness is induced by planetary perturbations, in particular 
     Jupiter's, and suggests that some of the recent, most powerful Earth 
     impacts may be associated with resonant groups of NEOs and/or very 
     young meteoroid streams. An intriguing consequence of this scenario is 
     that the impact hazard of Chelyabinsk-like objects should peak at 
     certain times in the year. 
  \end{abstract}


  \section{Introduction}
     Small bodies dominate the risk of sudden Earth impacts with just local, not global effects (Brown et al. 2013). Earth-striking 
     meteoroids are believed to be random in nature. In order to experience an impact, the nodal distance between the orbits of the Earth 
     and the parent body of the meteor must be smaller than the cross section of our planet for that particular body, and both have to pass 
     at the same time by the mutual node. If the nodes of the orbits of the impactors have a random and uniform distribution, any associated 
     impact events should be truly random. The paths followed by these objects are strongly perturbed, and exhibit fast and chaotic nodal 
     precession over time-scales of $\sim$10 Myr (see e.g. Ito \& Malhotra 2006). It is therefore not surprising that most studies assume 
     that the angular elements, in particular the longitude of the ascending node, of the orbits of near-Earth objects (NEOs) are randomly 
     and uniformly distributed in the range 0--2${\rm \pi}$. For many, this intrinsic chaoticity necessarily means that they are completely 
     random, and that all the impacts must be interpreted as uncorrelated events distributed according to Poisson statistics. But the 
     angular elements of the orbits of NEOs do not strictly follow uniform random distributions and this is the result of planetary 
     perturbations, mainly those of Jupiter (JeongAhn \& Malhotra 2014). 

     Even if the amount of data on powerful airburst events is still scarce and incomplete (Brown et al. 2013), the theoretical notion that 
     some of them may not be random is particularly relevant in the context of Planetary Defense initiatives and deserves to be explored. 
     However, and because of the small size of the current data set, the results of any such study shall be necessarily limited in scope as 
     only very large effects are likely to be reliably identified. Here, we use a time- and yield-limited sample of Earth-impacting 
     superbolides detected by Comprehensive Nuclear-Test-Ban Treaty Organization (CTBTO) infrasound sensors, and others from the literature, 
     since 2000\footnote{https://b612foundation.org/list-of-impacts-from-impact-video/}$^{,}$\footnote{http://newsroom.ctbto.org/2014/04/24/ctbto-detected-26-major-asteroid-impacts-in-earths-atmosphere-since-2000/} 
     to perform a rigorous statistical assessment of the existence of time-correlated impacts. The superbolide detections were reported by 
     the B612 Foundation using CTBTO data. This Letter is organized as follows. The sample used in this study is presented and discussed in 
     Section 2. Its presupposed randomness in time is put to the test in Section 3. Section 4 investigates the putative randomness in impact 
     coordinates. Possible sources of non-randomness are introduced in Section 5. Our results are discussed in Section 6 and conclusions are 
     summarized in Section 7. 

  \section{The B612 sample and more}
\begin{figure}
       \centering
        \includegraphics[width=\linewidth]{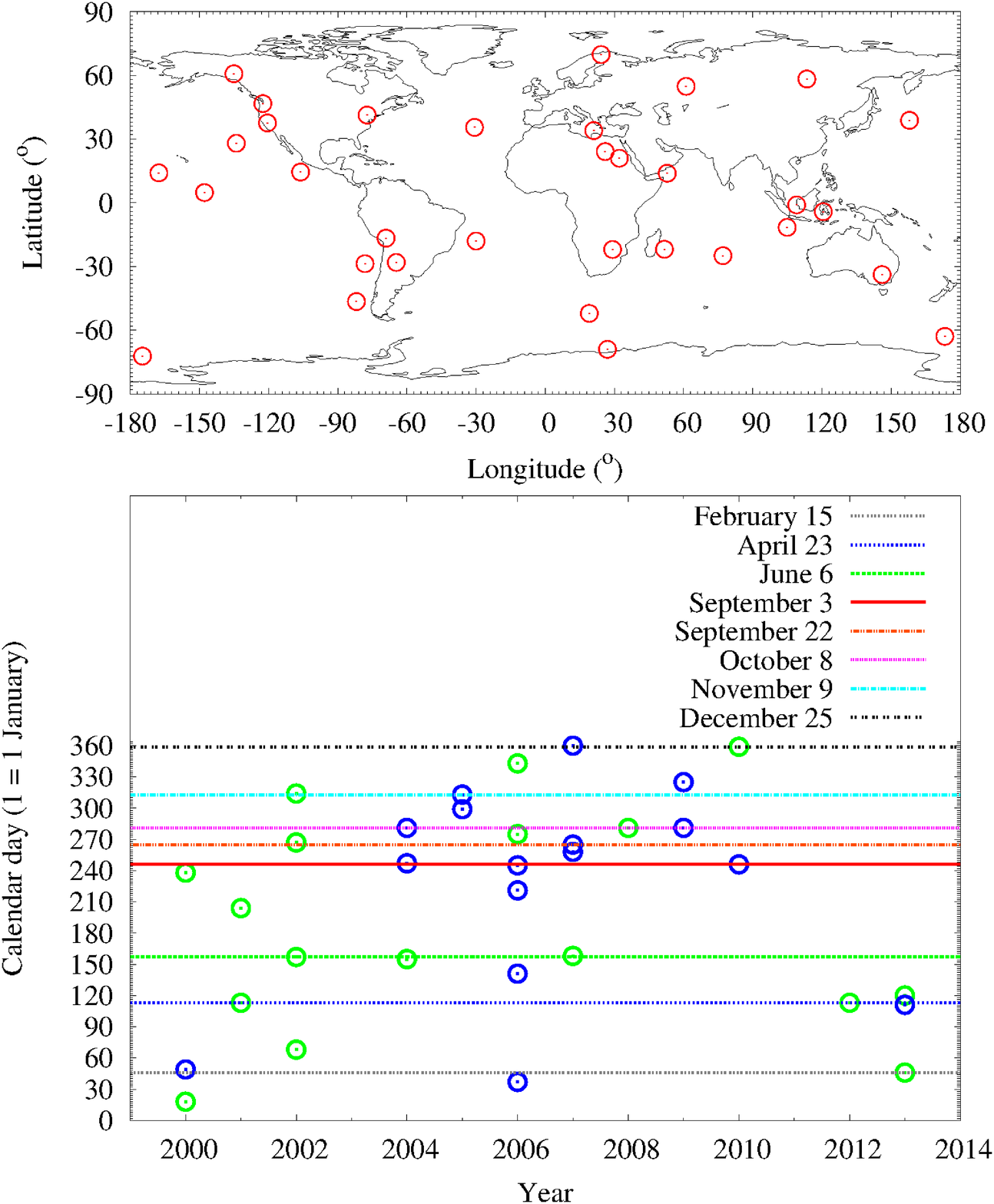}
        \caption{Geographical (top panel) and calendar-day (bottom panel) distributions of the events in Table \ref{events}. In the year-day
                 distribution diagram, green points correspond to impacts in the Northern hemisphere and blue points correspond to impacts 
                 in the Southern hemisphere. There are 17 impacts in the Northern hemisphere and 16 in the Southern one. The calendar-day
                 distribution for impacts in the Northern hemisphere is almost uniform. In contrast, the equivalent distribution for the
                 Southern hemisphere is very asymmetric. The overall impact risk for the Southern hemisphere is 50 per cent higher than that 
                 of the Northern hemisphere from September to December.
                }
        \label{sample}
     \end{figure}

     Since 2000, the CTBTO infrasound sensors of the International Monitoring System (IMS) network have detected 26 events with an 
     individual explosive energy in excess of 1 (and up to $\sim$500) kt of Trinitrotoluene (TNT) equivalent.$^{1}$ These superbolide 
     detections were reported by the B612 Foundation using CTBTO data; they are not the only detections compiled during that period of time. 
     Some of them are among the most energetic events ever instrumentally recorded; their yields are comparable to those of typical nuclear 
     weapons currently stocked. Higher yield, or amount of energy released, implies higher potential number of casualties and increased net 
     damage efficiency. In absence of clandestine nuclear tests, these events have been interpreted as resulting from Earth-impacting 
     fireballs or superbolides disintegrating as they travel through the atmosphere. The most extraordinary event recorded so far by the IMS 
     network is the Chelyabinsk superbolide, on 2013 February 15 (Brown et al. 2013; Le Pichon et al. 2013). The list of events appears in 
     Table \ref{events}, and their geographical and calendar day distributions are displayed in Fig. \ref{sample}. It includes seven 
     additional events from the published literature. They correspond to a time- and yield-limited sample of Earth-impacting superbolides. 

     The list of events in Table \ref{events} is controversial. It was released by the B612 Foundation on 2014 April 22 during a press 
     conference and it is based on research presented in Brown (2014). The original release included some errors. An 18 kt event 
     observed in Botswana on 2009 November 21\footnote{http://neo.jpl.nasa.gov/fireballs/} and six others with yields in the range 1.7--5 kt 
     were omitted. There was no Tasman Sea event on 2010 December 25, the latitude of the actual 33 kt event was 38\degr~N not S (see 
     footnote 3). The event listed as Indian Ocean on 2007 September 22 actually took place in the South Pacific Ocean. The event observed 
     from Finland on 2007 July 6 was not included in the video release. Our final list includes 33 events; as a time- and yield-limited 
     sample, it is probably quite complete. The size of this sample already suggests that the frequency of this type of impacts is from 3 to 
     10 times greater than previously believed (see original estimates in Brown et al. 2002). Any significant deviations from 
     near-completeness imply a further enhanced risk of locally dangerous impacts; for example, if the sample is just 50 per cent complete, 
     we should expect an average close to five such impacts per year instead of nearly two (see Fig. \ref{sample}, bottom panel). 

     Only in one case, the Almahatta Sitta event of 2008 in Sudan (Jenniskens et al. 2009), the incoming body was detected in advance but 
     only hours before impact. The first obvious conclusion to be drawn from these data is that asteroid impacts, at least those of objects 
     of a size under $\sim$20 m, are not extraordinarily rare. In addition, the fraction of our planet covered by oceans is 65.7 per cent 
     and the actual number of impacts away from land masses is 60.6 per cent; overall, 17 impacts are located in the Northern hemisphere. 
     This suggests a uniform geographical distribution of impacts, perhaps compatible with a random origin. However, if Earth-impacting 
     meteors are sporadic, random events, they are expected to be uniformly distributed in calendar day across the year. Thirteen impacts 
     have been recorded on the first part of the year and 20 on the second, a 21 per cent relative difference with respect to the 
     evenly-distributed scenario. This suggests a non-uniform temporal pattern of impacts. These apparent trends could be just coincidences 
     but, statistically speaking, what are the odds of observing patterns like the ones found in calendar day and geographical distribution?
%
%
%
%

  \section{Earth impacts: a probabilistic approach}
     The answer to the time-related section of the question asked above is connected with the famous "Birthday problem" of Probability 
     Theory (von Mises 1939; Abramson \& Moser 1970; Diaconis \& Mosteller 1989), the DNS cache poisoning technique used in Internet 
     attacks\footnote{http://www.secureworks.com/resources/articles/other\_articles/dns-cache-poisoning/} or DNA matching in Forensic 
     Science (Weir 2007): the probability of collision among a set of $n$ uniformly random numbers. In this context and if the nodes of the 
     orbits of NEOs are not uniformly distributed (see figs 4 and 5 in JeongAhn \& Malhotra 2014), a measurable deviation from a uniform 
     distribution in calendar day should be present in the available data.

     Let us consider a sample of $n$ Earth-impact events producing superbolides. We want to compute the probability of having two or more
     events on the same calendar day, but probably in different years, under the assumption that they are not time correlated. All calendar
     days have an equal chance of having an impact. We will count calendar days taking into account the existence of common and leap years 
     but, for the actual calculations, we will use years of 365 d. In our analysis, we will also study almost coincidences, when two or more 
     events are separated by one, two or three days. There are well-known formulae (e.g. Diaconis \& Mosteller 1989) for same calendar-day 
     events but, for practical reasons, the probabilities discussed here have been evaluated using Monte Carlo techniques as these are more 
     flexible. In Table \ref{events}, there are four same-day pairs (including a set of three events on the same calendar day, October 8). 
     For a sample of $n$ = 33 events, or 528 pairs, the probability of finding four pairs with the same calendar day is 0.0405. For such 
     sample and in the purely non-correlated case, the most probable number of pairs with the same calendar day is 1 (35.7 per cent). In 
     Table \ref{events}, there are nine pairs of events within one or less calendar-days difference. The likelihood of finding nine such 
     pairs in a non-correlated sample is 0.0190. There are also 14 pairs of events within two or less calendar-days difference. The 
     probability of finding such number of pairs in a non-correlated sample is 0.0093. Finally, there are 16 pairs of events within three or 
     less calendar-days difference but the likelihood of finding this many pairs in a non-correlated sample is 0.0216. The number of 
     same-day (or nearly same-day) coincidences is simply too high to be the result of chance alone. It is statistically obvious that the 
     impact events in Table \ref{events} are not uniformly distributed in time. They are far from sporadic in strict sense.

     The evaluation of the various probabilities presented here was carried out using the Monte Carlo method (Metropolis \& Ulam 1949). A 
     typical experiment consists of 10$^{10}$ tests of uniformly distributed (in the interval 1--365) integer samples of 33 numbers. For 
     each sample, the number of same (or near-same) calendar-day coincidences (or collisions) are counted and this value is divided by the 
     number of trials. Multiple experiments were performed to check for consistency and, when applicable, the probability values were 
     systematically checked against analytical results previously published (Abramson \& Moser 1970; Diaconis \& Mosteller 1989).

  \section{Geographical distribution}
\begin{figure}
       \centering
        \includegraphics[width=\linewidth]{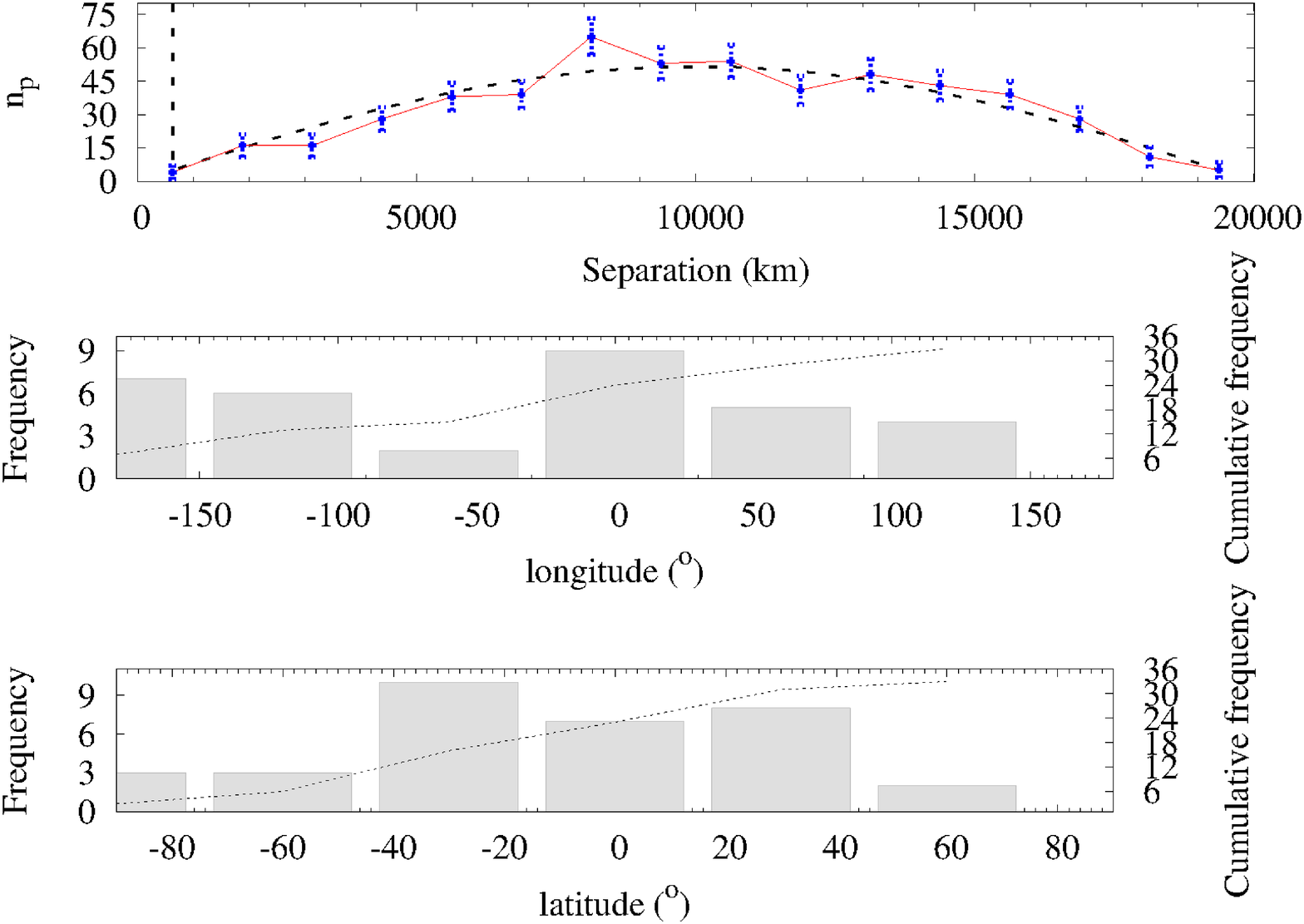}
        \caption{Separation distribution for the events in Table \ref{events} (top panel, red line and blue points with error bars), the 
                 number of pairs is $n_{\rm p}$. The theoretical separation distribution for an uncorrelated, uniformly distributed sample 
                 across the surface of a sphere of $n$ = 33 events (or 528 pairs) is displayed as a black discontinuous line. The number of 
                 bins is 2 $n_{\rm p}^{1/3}$. Distribution in longitude (middle panel) and latitude (bottom panel) for the same data. 
                }
        \label{dis}
     \end{figure}

     The answer to the geography-related section of the question is linked to the mathematical problem of "Sphere Point 
     Picking".\footnote{http://mathworld.wolfram.com/SpherePointPicking.html} In this section, the random points on the surface of a sphere 
     have been generated using an algorithm due to Marsaglia (1972). The continuous curve in Fig. \ref{dis} (top panel) is based on 
     experiments consisting of 10$^{6}$ tests. In any case, currently accepted models predict a non-uniform geographical distribution of 
     impacts (Le Feuvre \& Wieczorek 2008); as the source of the impactors (the NEO population) is not isotropic, latitudinal variations are 
     expected (see figs 6 and 7 in Le Feuvre \& Wieczorek 2008).
     
     Figure \ref{dis} (top panel) shows the expected distribution of separations between events on the surface of a sphere, black 
     discontinuous line, if they are uniformly distributed. The most probable separation is ${\rm \pi} r / 2$, where $r$ is the radius of the 
     sphere. The distribution associated with the data in Table \ref{events} is far from regular but we cannot conclude that the points are 
     not coming from a uniform distribution. Here, we adopt Poisson statistics and use the approximation given by Gehrels (1986) when 
     $n_{\rm p} < 21$: $\sigma \sim 1 + \sqrt{0.75 + n_{\rm p}}$, where $n_{\rm p}$ is the number of pairs. Only one bin shows a deviation 
     $>2\sigma$ (2.2$\sigma$). The distribution is similar for pairs of events within three or less calendar-days difference. The number of 
     impacts with latitude $\in(-40, 40)$\degr is 25 and outside that range we found 8 (see Fig. \ref{dis}, bottom panel); a uniform sample
     should have 21.2 and 11.8 events, respectively. In principle, this is consistent with predictions in Le Feuvre \& Wieczorek (2008). In 
     longitude, the distribution appears to be uniform (see Fig. \ref{dis}, middle panel) but there is a relatively large number of events 
     at longitudes in the range (15, 40)\degr E, 7. A uniform distribution predicts two events in that region, the deviation is close to 
     2$\sigma$. 

     On the other hand, the calendar-day distribution for impact events in the Northern hemisphere is virtually uniform with nine impacts in 
     the first half of the year and eight during the second half. Surprisingly, the equivalent distribution for the Southern hemisphere is 
     markedly asymmetric with 4 events in the first half of the year and 12 during the second half. This is a 2$\sigma$ departure from an 
     isotropic distribution, where $\sigma$ is now the standard deviation for binomial statistics. Even if modest, this asymmetry cannot be 
     the result of the relative orientation because for impacts in the Southern hemisphere the angle of the ecliptic is shallowest near the 
     autumnal equinox in September. In fact, the observed annual variation of sporadic meteor rates is inconsistent with the pattern 
     followed by the impact events in Table \ref{events}. For the Northern hemisphere, the highest rate is observed in November and the 
     lowest in May--June; for the Southern hemisphere, the highest is observed in June and the lowest in September--October (Jenniskens 
     1994; Poole 1997; Campbell-Brown \& Jones 2006). Sporadic meteor activity appears to be higher when the ecliptic is at its highest on 
     the sky. These facts strongly suggest that a fraction of the most recent multi-kiloton impacts are not associated with the usual 
     sporadic meteor population. However, a somewhat similar pattern is observed in the analysis of 259 fireballs carried out by Halliday,
     Griffin \& Blackwell (1996). Therefore, there is a genuine excess of relatively large objects with nodes in the range of values 
     appropriate to produce an impact on Earth from September to December. For this type of objects, the overall impact risk for the 
     Southern hemisphere is 50 per cent higher than that of the Northern hemisphere within the same time frame. Assuming that the shape of 
     the Earth is not responsible for this asymmetry, the apparent excess must be produced by some type of orbitally coherent but discrete 
     sources.
%
%
%
%

  \section{Young streams versus resonant groups}
     The Chelyabinsk event of 2013 February 15 and the South China Sea event of 2000 February 18 are separated by three calendar days and 
     their impact nodes are close to 146\fdg5. Both impacts may have a common source. The pre-impact orbit of the parent body of the 
     Chelyabinsk impactor is well understood (see e.g. de la Fuente Marcos \& de la Fuente Marcos 2014) and it may be associated with a 
     resonant family of asteroids. The North Pacific Ocean event of 2001 April 23, between Hawaii and California, and the Sutter's Mill 
     meteor of 2012 April 22 were observed on the same calendar day. The Santiago del Estero event in Argentina was observed on 2013 April 
     21. The impacting node was $\sim$212\fdg5. For the Sutter's Mill meteor, it happened at the descending node; the parent body being an 
     object moving in an orbit close to those of the Jupiter-family comets (Jenniskens et al. 2012). The two same-day impacts are very 
     likely related; the Argentinian event appears to be unrelated, its properties being too different (see footnote 3). The Crete bolide of
     2002 June 6 (Brown et al. 2002) and the Reisadalen event of 2007 June 7, close to the border between Norway and Finland, were observed 
     within one calendar day. The impact node was nearly 255\fdg5 and both happened at the descending node. The bright fireball observed 
     over Washington State on 2004 June 3 may be related to them. Within one calendar day, we have the Southern Ocean event of 2004 
     September 3 near Antarctica, the Indian Ocean event of 2006 September 2, and the South Pacific Ocean event of 2010 September 3. These 
     three events exhibit the most robust signs of time correlation with impact at the descending node $\sim$340\degr. It is thought that 
     the parent body of the Southern Ocean event was a small Aten asteroid with an orbital period of 293 d (Klekociuk et al. 2005). Atens 
     have the shortest lifetimes against collision with our planet (e.g. Chyba 1993; Steel 1995; Dvorak \& Freistetter 2001). The South 
     Pacific Ocean event of 2007 September 22 and the Vitim or Bodaybo event of 2002 September 24 are separated by two calendar days and 
     their impact nodes are close to 359\degr. On the same calendar day we find the Indian Ocean event of 2004 October 7, the Almahatta 
     Sitta event of 2008 October 7 in Sudan, and the South Sulawesi event of 2009 October 8 in Indonesia. An analysis of the position of the 
     impact node on those dates ($\sim$14\degr) shows that the Indian Ocean and the Almahatta Sitta events are likely related. The South 
     Sulawesi and Almahatta Sitta events are almost certainly unrelated (see footnote 3). The Almahatta Sitta event was caused by the impact 
     of the Apollo asteroid 2008 TC$_{3}$ (Jenniskens et al. 2009). The origin of asteroid 2008 TC$_{3}$ was traced to the Nysa-Polana 
     asteroid family (Gayon-Markt et al. 2012). On 2002 November 10, there was an impact event on the North Pacific Ocean and on 2005 
     November 9 another impact was observed in New South Wales, Australia. The impact node was nearly 47$^{o}$. The South Pacific Ocean 
     event of 2007 December 26 and the North Pacific Ocean event of 2010 December 25 have an impact node $\sim$93\degr. 

     More than 50 per cent (20/33) of the events in Table \ref{events} show some degree of time correlation and the analysis of the 
     geographical distribution also suggests that a certain degree of spatial correlation at the $\sim$2$\sigma$ level could be present, 
     see Fig. \ref{dis}. For decades, it has been assumed that Earth-impact events were uncorrelated. Unlike prior efforts, our statistical 
     analysis shows that the customary assumption that Earth-impact events are sporadic and of random origin is incorrect, at least in the 
     case of the most powerful and recent ones. The most natural and simple explanation is to admit the existence of very young meteoroid 
     streams, resulting from the breakup of an asteroid or comet, and/or resonant streams in which secular resonances force a group of 
     objects to orbit around the Sun following similar orbits; impacts occur when their paths intersect that of our planet. A resonant 
     stream may well be the source of the parent body of the Chelyabinsk superbolide and related bodies (de la Fuente Marcos \& de la Fuente 
     Marcos 2013, 2014). Very young meteoroid streams may be disrupted in just 30 years or less (Lai et al. 2014); resonant streams are 
     somewhat permanent when compared with their ephemeral relatives resulting from breakups because they are continually being replenished 
     via resonance capture. Hybrid streams may exist (see e. g. Williams \& Ryabova 2011). The material trapped in a resonant stream may be 
     diverse in composition producing meteorites of various types but debris from a breakup is expected to be homogeneous in composition. 
     
  \section{Discussion}
     Although our analysis provides robust statistical proof that the cadence of recent multi-kiloton impact events is incompatible with a 
     random fall pattern at the 0.05 significance level or better, it may be argued that the data in Table \ref{events} are not an accurate
     temporal and spatial representation of all superbolides striking Earth with yields $>$ 1 kt within the time frame 2000-2013. It is 
     true that for the events included here the uncertainties in impact time ($\sim$10 s) and coordinates ($\sim$1\degr) are too small to 
     affect significantly any of our conclusions, but yields are model dependent and a difference of a factor of 3 between models is not 
     unusual (Brown et al. 2013). However, this is only relevant for yields $<$ 3 kt. On the other hand, a number of events with explosive 
     energy under 1 kt, but within the factor 3 region, have been left outside our analysis; their number may compensate for any wrongfully 
     included, underperforming event. As pointed out above, we believe that the sample used here is reasonably complete, especially for the 
     most powerful events. An average annual number of multi-kiloton impacts much higher than three does not appear to be supported by 
     currently available observations. It is true that the geographical detection efficiency during the first half of the period studied 
     here, when the IMS network was far from complete, was lower (Le Pichon, Ceranna \& Vergoz 2012). The IMS network was 75 per cent 
     complete by the end of 2013; now it is almost 80 per cent complete. Seasonal variations in the stratospheric winds are also critical, 
     IMS detection capabilities are best around January and July, and worst near the equinox periods (Le Pichon et al. 2012). In their 
     study, Le Pichon et al. (2012) consider ground sources and not superbolide outbursts above the tropopause; their results are based on 
     the full IMS infrasound network, not the currently installed network. From Table \ref{events}, we have that 20 events were observed 
     from 2000 to 2006 and 13 events from 2007 to 2013. Also, only one event was observed in January, July and March; September has the 
     record with six detections. In summary, no statistically significant biases of systematic nature appear to be present in our sample. 
     The year-to-year variation in the number of impacts can be naturally explained within the context of secular resonances (see the 
     clustered, periodic close-approaches characteristic of resonant groups in fig. 4 in de la Fuente Marcos \& de la Fuente Marcos 2014). 
     The only true bias affects bodies moving in very Earth-like orbits; if the orbit of a meteoroid is only slightly different than that of 
     the Earth, the encounter speed in a collision may be very low and slower meteors produce less ionization. Such impactors are much less 
     likely to be observed (using either radar, infrasound or optical equipment) than higher encounter speed meteors moving in very 
     eccentric orbits, unless they are relatively large. No objects of this type appear to be included in Table \ref{events}.
    
     Regarding Earth strikes, a coherent debris stream, resonant or not, is characterized by its radiant. If an impactor is associated with
     a certain stream, its radiant must be above the local horizon at the time of the event. Although a putative radiant above the horizon 
     at the time of impact does not secure a linkage, the opposite is true: if the radiant is below the horizon, we can strictly rule out 
     any connection. This criterion has been used to discard the 2013 April 21 or the 2009 October 8 events in the previous section. It is 
     widely accepted that orbital destruction is far too efficient to allow the existence of long-lived near-Earth meteoroid streams. 
     Near-Earth meteoroid streams are made of debris of asteroidal or cometary origin (Jopek \& Williams 2013). Groups of objects moving 
     initially in similar trajectories lose all orbital coherence in a short time-scale (Pauls \& Gladman 2005; Rubin \& Matson 2008; Lai et 
     al. 2014). However, small bodies part of debris streams formed during the last few decades may still follow very similar orbits,
     including having a common node. Such concentration of nodes has not been observed for large NEOs. Schunov\'a et al. (2012) could not 
     find any significant near-Earth meteoroid streams among currently known objects but they could not refute their existence. However, 
     Schunov\'a et al. (2014) confirmed that streams from tidally disrupted objects can only be detected for a few thousand years after the 
     disruption event and only if the parent body is large enough. The debris field made of 10$^5$ 1--10 m fragments of a disrupted 
     progenitor smaller than about 350 m in diameter cannot be detected by current techniques. Such a very young meteoroid stream could 
     easily be the source of some of the events discussed here. An example of an already-evolved meteoroid stream is the one responsible for
     the Geminids meteor shower. The parent body of this stream is the asteroid (3200) Phaethon (e.g. de Le\'on et al. 2010) but other 
     smaller bodies like (155140) 2005 UD (Kinoshita et al. 2007) or (225416) 1999 YC (Kasuga \& Jewitt 2008) are very probably associated 
     fragments. Phaethon is still actively producing meteoroids (Jewitt \& Li 2010; Ryabova 2012; Li \& Jewitt 2013) that may eventually end 
     up as meteorites (Madiedo et al. 2013).

  \section{Conclusions}
     The pattern of recent multi-kiloton impact events is not random. This conclusion cannot be attributed to observational biases and may
     have its origin in secular planetary perturbations. Our main results are summarized as follows:
     \begin{itemize}
        \item The cadence of recent multi-kiloton impact events is incompatible with a random fall pattern at the 0.05 significance level or 
              better.
        \item Statistically, the impact coordinates of these events do not follow a uniform distribution at the $\sim$2$\sigma$ level.
        \item Non-random Earth impacts may have their source in very young and/or resonant streams, eight candidates are proposed.
     \end{itemize}
     Resonant and/or very young meteoroid streams may dominate the flux of impactors having just local effects. The intrinsic chaoticity 
     characteristic of these streams makes long-term predictions relatively useless in this case and forces any realistic Planetary Defense 
     programme to keep an eye permanently on this population; quite literally, the meteoroid streams responsible for local impacts today may 
     not be the same a few decades from now. Although most of the impact events in Table \ref{events} ended up with the disintegration of 
     the incoming body high up in the atmosphere, sometimes above remote parts of the ocean, causing virtually no problems on the ground 
     (Chelyabinsk event excluded, Popova et al. 2013), early identification of time- and, perhaps, space-correlated events may help 
     mitigating subsequent impacts associated with the same streams.

  \section*{Acknowledgements}
     We would like to thank the referee, B. Ivanov, for his helpful and constructive report, and P. Brown and T. Muetzelburg for answering 
     questions regarding the IMS network. P. Mialle provided an additional review from the part of the CTBTO. This work was partially 
     supported by the Spanish `Comunidad de Madrid' under grant CAM S2009/ESP-1496. We thank M. J. Fern\'andez-Figueroa, M. Rego Fern\'andez 
     and the Department of Astrophysics of the Universidad Complutense de Madrid (UCM) for providing computing facilities. Most of the 
     calculations and part of the data analysis were completed on the UCM's `Servidor Central de C\'alculo' and we thank the computing staff 
     (S. Cano Alsua) for help during this stage. In preparation of this Letter, we made use of the NASA Astrophysics Data System and the 
     ASTRO-PH e-print server.

  \newpage
  \appendix
  \section{The time- and yield-limited sample of Earth-impacting superbolides}
     As pointed out above, the time- and yield-limited sample of Earth-impacting superbolides used in this research is based on the 
     controversial B612 sample (26 events) and includes seven additional events from the literature. The list of events appears in Table 
     \ref{events}.   
%
%
     \begin{table}
      \centering
      \fontsize{8}{11pt}\selectfont
      \tabcolsep 0.35truecm
      \caption{The time- and yield-limited sample of Earth-impacting superbolides used in this research.
              }
      \begin{tabular}{ccc}
       \hline
          Date       & Location              & Calendar day \\
       \hline
          01/18/2000 & Tagish Lake           &  18          \\ 
          02/06/2006 & South Atlantic Ocean  &  37          \\
          02/15/2013 & Chelyabinsk Oblast    &  46          \\
          02/18/2000 & South China Sea       &  49          \\
          03/09/2002 & North Pacific Ocean   &  68          \\
          04/21/2013 & Santiago del Estero   & 111          \\
          04/22/2012 & California            & 113          \\
          04/23/2001 & North Pacific Ocean   & 113          \\
          04/30/2013 & North Atlantic Ocean  & 120          \\
          05/21/2006 & South Atlantic Ocean  & 141          \\
          06/03/2004 & Washington State      & 155          \\
          06/06/2002 & Mediterranean Sea     & 157          \\
          06/07/2007 & Finland               & 158          \\
          07/23/2001 & Pennsylvania          & 204          \\
          08/09/2006 & Indian Ocean          & 221          \\
          08/25/2000 & North Pacific Ocean   & 238          \\
          09/02/2006 & Indian Ocean          & 245          \\
          09/03/2010 & South Pacific Ocean   & 246          \\
          09/03/2004 & Southern Ocean        & 247          \\
          09/15/2007 & Carancas              & 258          \\
          09/22/2007 & South Pacific Ocean   & 265          \\
          09/24/2002 & Vitim, Siberia        & 267          \\
          10/02/2006 & Arabian Sea           & 275          \\
          10/07/2004 & Indian Ocean          & 281          \\
          10/07/2008 & Sudan                 & 281          \\
          10/08/2009 & South Sulawesi        & 281          \\
          10/26/2005 & South Pacific Ocean   & 299          \\
          11/09/2005 & New South Wales       & 313          \\
          11/10/2002 & North Pacific Ocean   & 314          \\
          11/21/2009 & Botswana              & 325          \\
          12/09/2006 & Egypt                 & 343          \\
          12/25/2010 & North Pacific Ocean   & 359          \\
          12/26/2007 & South Pacific Ocean   & 360          \\
       \hline
      \end{tabular}
      \label{events}
     \end{table}
%
%

  \section{NOTE ADDED IN PROOF}
     After this work was accepted by MNRAS Letters, J. M. Madiedo pointed out that the following paper had arrived at similar conclusions:

     S\'anchez de Miguel A., Oca\~na F., Zamora S., Tapia C., Santamaria A., de Burgos A., 2014,
     in Garz\'on Guerrero J. A., L\'opez S\'anchez A. R., eds,
     Libro de Actas del XXI Congreso Estatal de Astronom\'{\i}a.
     Granada, Spain

     However, we remark that the sample used in the study of S\'anchez de Miguel et al. (2014) presents multiple biases: notably, it is 
     sparse in time and locally restricted, to the USA and Spain. Furthermore, its primary sources (newspapers) are not necessarily 
     reliable.
\end{document}